\documentclass[aps,prd,
preprintnumbers,groupedaddress]{revtex4}
\usepackage{epsbox}
\usepackage{amsmath}
\usepackage[dvips]{graphicx}

\begin{document}
\preprint{\vtop{
{\hbox{YITP-11-22}\vskip-0pt
                 \hbox{KANAZAWA-11-03} \vskip-0pt
}
}
}


\title{ 
Tetra-quark mesons with exotic quantum numbers\footnote{
Talk given at Baryons'10, International Conference on the Structure of Baryons, 
Dec. 7 -- 11, 2010, Osaka, Japan} 
\\
-- {\it Their production and related} --}

\author{
Kunihiko Terasaki   
}
\affiliation{
Yukawa Institute for Theoretical Physics, Kyoto University,
Kyoto 606-8502, Japan \\
Institute for Theoretical Physics, Kanazawa University, 
Kanazawa 920-1192, Japan
}

\begin{abstract}
{Tetra-quark mesons with exotic quantum numbers and their production rates are 
studied. }
\end{abstract}

\maketitle

Tetra-quark mesons can be classified into the following four groups~\cite{Jaffe} 
(and \cite{D_{s0}-KT}),  
\begin{eqnarray} 
&&\hspace{-8mm} \{qq\bar q\bar q\} =  
[qq][\bar q\bar q] \oplus (qq)(\bar q\bar q)  
\oplus \{[qq](\bar q\bar q)\oplus (qq)[\bar q\bar q]\} 
                                                   \label{eq:4-quark} 
\end{eqnarray} 
with $q = u,\,d,\,s$ (and $c$), where parentheses and square brackets denote 
symmetry and anti-symmetry, respectively, of flavor wavefunctions (wfs.) under 
exchange of flavors between them. 
Each term in the right-hand-side of Eq.~(\ref{eq:4-quark}) is again classified into 
two groups with  
${\bf \bar 3_c}\times{\bf 3_c}$ and ${\bf 6_c}\times {\bf \bar 6_c}$  
of the color $SU_c(3)$~\cite{Jaffe}. 
In the case of heavy mesons, the ${\bf \bar 3_c}\times{\bf 3_c}$ state is taken 
as the (dominant part of) lower lying one ~\cite{KT-Hadron-2003,HT-isospin}. 
Then, spin ($J$) and parity ($P$) of corresponding $[qq][\bar q\bar q]$ and 
$[qq](\bar q\bar q)\pm (qq)[\bar q\bar q]$ mesons are $J^P=0^+$ and $1^+$, 
respectively.      
However, we ignore $(qq)(\bar q\bar q)$~\cite{NFQCD-KT}, although still 
controversial~~\cite{K-pi-3/2}. 

One of candidates of heavy tetra-quark mesons is  $D_{s0}^+(2317)$. 
It was discovered in the $D_s^+\pi^0$ mass distribution, while no signal in the 
$D_s^{*+}\gamma$ channel in inclusive $e^+e^-$ 
annihilation~\cite{Babar-e^+e^-D_{s0},CLEO-D_{s0}}. 
This fact suggests that the decay $D_{s0}^+\rightarrow D_s^+\pi^0$ is dominant and 
caused by the isospin conserving strong interaction, because of the well-known 
hierarchy of hadron interactions~\cite{HT-isospin}, 
$|${\it isospin conserving strong ones}$|$ $\gg$ $|${\it electromagnetic ones}$|$ $\gg$ 
$|${\it isospin non-conserving ones}$|$. 
Here the last one is of the order of $\alpha$~\cite{Dalitz}, where $\alpha$ is the 
fine structure constant. 
Therefore, $D_{s0}^+(2317)$ should be an iso-triplet state, and hence it is natural to 
assign $D_{s0}^+(2317)$ to 
$\hat F_I^+\sim [cn][\bar s\bar n]_{I=1},\,(n=u,\,d)$~\cite{D_{s0}-KT}. 
In this case, the observed narrow width~\cite{PDG10} of $D_{s0}^+(2317)$ is understood 
by a small overlap of color and spin wfs.~\cite{KT-Hadron-2003,HT-isospin,ECT-talk}. 
The above assignment is consistent~\cite{NFQCD-KT,QNP06-KT,Scadron-70} with 
the observation in $B$ decays~\cite{Belle-D_{s0}},  
$Br(B\rightarrow\bar{{D}}
\tilde{{D}}_{{s0}}^{{+}}{(2317)}[{D_s}^{{+}}{\pi^0}]) = 
(8.5^{{+}2.1}_{{-}1.9} {\pm} 2.6)\times 10^{-4}$ and 
${Br}({B}\rightarrow \bar{{D}}\tilde{{D}}_{{s0}}^{{+}}{(2317)}
{[D_s^{{*+}}{\gamma}])}                                  
{\,\,=(2.5^{{+}2.0}_{{-}1.8}({<} 7.5))\times 10^{-4}}$, 
where signals observed in the $D_s^+\pi^0$ and $D_s^{*+}\gamma$ are denoted as 
$\tilde{D}_{s0}^{+}(2317)[D_s^{+}\pi^0]$ and $\tilde{D}_{s0}^{+}(2317)[D_s^{*+}\gamma]$, 
respectively. 
Therefore, $\hat F_I^+$ and its iso-singlet partner 
$\hat F_0^+\sim [cn][\bar s\bar n]_{I=0}$ are identified with 
$\tilde{D}_{s0}^{+}(2317)[D_s^{+}\pi^0]$ and  $\tilde{D}_{s0}^{+}(2317)[D_s^{*+}\gamma]$, 
respectively, because the former decays dominantly into the $D_s^+\pi^0$ state while 
the latter into the $D_s^{*+}\gamma$ due to the above hierarchy of hadron 
interactions.     
For more details, see Refs.~\cite{D_{s0}-KT}, \cite{HT-isospin}, \cite{NFQCD-KT} and 
\cite{ECT-talk}. 

Another candidate of tetra-quark meson is $X(3872)$ with $J^{P\mathcal{C}}=1^{++}$, 
where $\mathcal{C}$ is the charge conjugation parity. 
It was discovered~\cite{Belle-X(3872)} and confirmed~\cite{confirm-X} 
in the $\pi^+\pi^-J/\psi$ mass distribution. 
However, the $X(3872)\rightarrow\pi^+\pi^-J/\psi$ decay violates badly isospin 
conservation~\cite{Belle-X-gamma-psi,CDF-pipi} which works well in ordinary strong 
interactions. 
Such a large violation of isospin symmetry can be understood~\cite{omega-rho-KT} 
by the $\omega\rho^0$ mixing which is well-known as the origin of isospin 
non-conservation in nuclear forces~\cite{omega-rho}, because the $\rho^0$ pole 
contribution is enhanced in the $X(3872)\rightarrow\pi^+\pi^-J/\psi$ decay due to 
$|m_\omega - m_\rho| \ll |m_\omega|$. 
In fact, the measured values of the ratio of decay rates 
\begin{equation}
R^\gamma\equiv \frac{Br(X(3872)\rightarrow \gamma J/\psi)} 
{Br(X(3872)\rightarrow \pi^+\pi^- J/\psi)} = \left\{
\begin{tabular}{l}
$0.14 \pm 0.05$,    (Belle~\cite{Belle-X-gamma-psi} )\\
 $0.33 \pm 0.12$,   (Babar~\cite{Babar-gamma-psi'})
 \end{tabular}\right.
\end{equation} 
have been approximately reproduced, i.e., 
$(R^\gamma)_{{\rm tetra}}\simeq  (R^\gamma)_{\rm Babar} 
\simeq (R^\gamma)_{\rm Belle}$~\cite{NFQCD-KT,omega-rho-KT},  
by assuming that  $X(3872)$ is a tetra-quark system like a 
$\{[cn](\bar c\bar n) + (cn)[\bar c\bar n]\}_{I=0}$ meson~\cite{Terasaki-X} 
(or a $D^0\bar D^{*0}$ molecule~\cite{Toernquist}) and that the isospin 
non-conservation under consideration is caused by the 
$\omega\rho^0$ mixing.   
In contrast, if $X(3872)$ were assumed to be a charmonium $X_{c\bar c}$ with 
$J^{P\mathcal{C}} = 1^{++}$, the above ratio could not be 
reproduced, i.e.,  
$(R^\gamma)_{{c\bar c}}\gg (R^\gamma)_{\rm Babar} 
\simeq (R^\gamma)_{\rm Belle}$~\cite{NFQCD-KT,omega-rho-KT}.  
Therefore, we have seen that a tetra-quark interpretation of $X(3872)$ is favored 
over the charmonium. 
In addition, production~\cite{CDF-prompt-X} of the {\it prompt} $X(3872)$ seems to 
favor a compact object like a tetra-quark meson over an extended object like a 
loosely bound molecule~\cite{compact-prompt-X}, and hence the tetra-quark model 
mentioned above survives while the $D^0\bar D^{*0}$ molecular model would be ruled 
out. 

Although quantum numbers of $D_{s0}^+(2317)$ and $X(3872)$ are not exotic, 
their tetra-quark interpretation has been favored by experiments as seen above, 
so that existence of their partners with exotic quantum numbers is expected. 
However, for example, neutral and doubly charged partners, $\hat F_I^0$ and 
$\hat F_I^{++}$, of $\hat F_I^+ = D_{so}^+(2317)$ have not been observed in inclusive 
$e^+e^-$ annihilation~\cite{Babar-D_{s0}-charged-partners}. 
Nevertheless, it does not necessarily imply their non-existence but it might suggest  
that their production is suppressed in this process.
This can be understood by considering their production in the framework of 
minimal $q\bar q$ pair creation~\cite{Scadron-70}.  
On the other hand, their production rates in $B$ decays have been very crudely 
estimated as 
$Br(B_u^+\rightarrow D^-\hat F_I^{++})
\sim Br(B_d^0\rightarrow \bar D^0\hat F_I^{0})  
\sim Br(B_u^+(B_d^0)\rightarrow \bar D^0(D^-)\hat F_0^+)         
\sim Br(B_u^+(B_d^0)\rightarrow \bar D^0(D^-)\hat F_I^+)_{\rm exp}  
\sim 10^{-(4 - 3)}$,     
because the above decays are described by the same type of quark-line diagrams 
and hence the sizes of their amplitudes are expected to be nearly equal to each 
other~\cite{QNP06-KT,Scadron-70,production-D_{s0}-KT}. 

Observations of mesons with exotic quantum number(s) provide additional evidences 
for existence of tetra-quark mesons. 
In our scheme,  $\hat E^0\sim [cs][\bar u\bar d]$ meson~\cite{D_{s0}-KT} is only one 
scalar meson with ${C} = -S = +1$. 
Axial-vector mesons with exotic quantum numbers, which come from 
$\{[qq](\bar q\bar q)\oplus (qq)[\bar q\bar q]\}$, are 
$H_{Acc}^+\sim (cc)[\bar u\bar d]$ with $C=2$, $S=0$, $I=0$; 
$K_{Acc}\sim (cc)[\bar n\bar s]$ with $C=2$, $S=1$, $I=1/2$; 
$E^0_{A(cs)}\sim (cs)[\bar u\bar d]$ with $C=1$, $S=-1$, $I=0$; 
$E_{A[cs]}\sim [cs](\bar n\bar n)$ with $C=1$, $S=-1$, $I=1$. 
Their masses can be very crudely estimated as 
$m_{\hat E^0}\simeq 2.32$ GeV, $m_{H_{Acc}}\simeq  3.87$ GeV, 
$m_{K_{Acc}}\simeq  3.97$ GeV, 
$m_{E_{A(cs)}}\simeq  m_{E_{A[cs]}}\simeq 2.97$ GeV 
by using a quark counting with $m_c - m_s\simeq 1.0$ GeV and $m_s - m_n\simeq 0.1$ 
GeV as in Ref.~\cite{hidden-charm-scalar-KT}, where $m_{D_{s0}(2317)} \simeq 2317$ 
MeV and $m_{X(3872)}\simeq 3872$ MeV have been taken as the input data. 
It should be noted that we have predicted~\cite{hidden-charm-scalar-KT} the mass of 
hidden-charm iso-triplet scalar $\hat\delta^c_{I=1}$ to be 
$m_{\hat\delta^c_{I=1}}\simeq 3.3$ GeV, using the same quark counting, and that 
the result fits much better to a peak at $3.2$ GeV in the $\eta\pi$ channel, which 
was observed by the Belle~\cite{hidden-charm-scalar-Belle} and can be considered as 
a signal of $\hat\delta^c_{I=1}$, than predictions of the corresponding meson mass by 
the other models~\cite{Maiani,hidden-charm-scalar}. 

Production of tetra-quark mesons is now in order. 
Productions of $K_{Acc}^+$ and $K_{Acc}^{++}$ can be described by the quark-line 
diagram, Fig.~1(c), which is of the same type as Fig.~2(a) and Fig.~3(b) in 
Ref.~\cite{production-D_{s0}-KT} describing 
$B_u^+\rightarrow \bar D^0\hat F_I^+$ and $B_d^0\rightarrow D^-\hat F_I^+$,  
respectively. 
Because 
$Br(B_u^+(B_d^0)\rightarrow \bar D^{0}(D^-)D_{s0}^+(2317))_{\rm exp}
\sim 10^{-(4 - 3)}$    
as mentioned before, production rates for $K_{Acc}^+$ and $K_{Acc}^{++}$ would be 
very crudely estimated as 
\begin{equation}Br(B_c^+\rightarrow D^{*-}K_{Acc}^{++})\sim 
Br(B_c^+\rightarrow \bar D^{*0}K_{Acc}^+)\sim 10^{-(4 - 3)},   \label{eq:rates-for-K_A}
\end{equation}  
because differences of kinematics between $B_n$ and $B_c$ decays under 
consideration do not change  order of magnitude of their branching 
fractions~\cite{exotic-quantum-numbers-KT}. 
Production of $H_{Acc}^+$ is described by the diagram Fig.~1(d) which describes 
the CKM suppressed decay, so that the rate for $H_{Acc}^+$ production would be 
more suppressed by a factor $\sim |V_{cd}/V_{cs}|^2\simeq 0.05$ than the above ones, 
where $V_{cd}$ and $V_{cs}$ are the CKM matrix elements, 
although it is described by the same type of diagram as the previous ones. 
Productions of scalar and axial-vector tetra-quark mesons with ${C} = -S = 1$ can be 
described by the diagrams, (a) and (b) in Fig.~1. 
These diagrams are of the same type as that of Fig.~4(c) in 
Ref.~\cite{production-D_{s0}-KT} describing $\bar B_d^0\rightarrow K^-\hat F_I^+$ 
whose rate has already been measured~\cite{BELLE-D_{s0}-K^-}.    
However, the result is smaller by about an order of magnitude than 
$Br(\bar B_d^0\rightarrow D^-D_{s0}^+(2317))_{\rm exp}$ because the former includes 
an $s\bar s$ pair creation, as discussed in Refs.~\cite{production-D_{s0}-KT} and 
\cite{exotic-quantum-numbers-KT}. 
In contrast, we now expect that rates for decays described by the diagrams (a) and (b) 
in Fig.~1 are not suppressed, because these diagrams involve no $s\bar s$ creation, i.e., 
\begin{eqnarray}
&&
Br( B_u^-\rightarrow D^{*-}E^0_{A(cs)})                 
\sim Br( B_u^-\rightarrow D^{*-}E^0_{A[cs]}) 
\sim Br( B_u^-\rightarrow D^-\hat E^0)       \nonumber\\
&&
\sim Br(\bar B_d^0\rightarrow \bar D^{*0}E^0_{A(cs)})
\sim Br(\bar B_d^0\rightarrow \bar D^{*0}E^0_{A[cs]})
\sim Br(\bar B_d^0\rightarrow \bar D^0\hat E^0) \nonumber\\
&&\sim 10\times 
Br(\bar B_d^0\rightarrow K^-D_{s0}^+(2317))\sim 10^{-(4-3)}.  
                                                                          \label{eq:rates-for-E}
\end{eqnarray}

Although decay properties of these exotic mesons would be useful to search for 
them, rates for two- and body-decays of, in particular,  $K_{A{cc}}$ and $H_{A{cc}}^+$ 
would be crucially sensitive to their mass values because they are estimated to be 
very close to their corresponding thresholds. 
Therefore, calculations of these rates would be keenly model dependent at 
the present stage. 
In addition, no experimental data which can be used  as the input data is known, 
so that they are left as one of our future subjects. 

\begin{figure}[b]    
\includegraphics[width=130mm,clip]{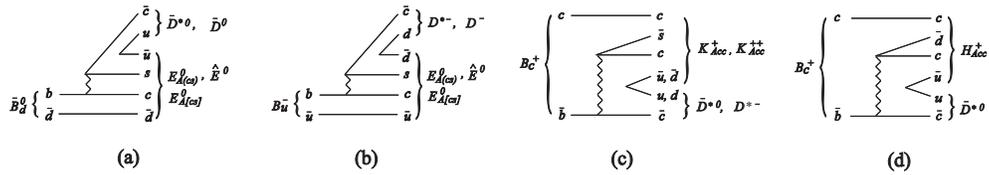}
\caption{Productions of tetra-quark scalar 
and axial-vector mesons with exotic quantum numbers. }   
\end{figure}   

In summary, we have studied scalar and axial-vector mesons with exotic quantum 
numbers, and have estimated their production rates, comparing quark-line diagrams 
describing their productions with those of $D_{s0}^+(2317)$. As the result, we have 
seen that a major part of them can be large enough to be observed in $B$ decays. 
\section*{Acknowledgments}    
The author would like to thank Professor H.~J.~Lipkin and Professor T.~Iijima for 
discussions and comments. 
He also would like to appreciate the organizers for financial support. 



\end{document}